\documentclass[10pt,twocolumn,letterpaper]{article}

\usepackage{iccv}
\usepackage{times}
\usepackage{epsfig}
\usepackage{graphicx}
\usepackage{amsmath}
\usepackage{amssymb}
\usepackage{adjustbox}

\newtheorem{theorem}{Theorem}[section]

\newtheorem{lemma}[theorem]{Lemma}
\newcommand{\mli}[1]{\mathit{#1}}

\newcommand{\mxy}[1]{\textcolor{black}{#1}}
\newcommand{\GL}[1]{\textcolor{black}{#1}}

\usepackage[breaklinks=true,bookmarks=false]{hyperref}

\iccvfinalcopy 

\def\iccvPaperID{1057} 

\ificcvfinal\fi

\begin{document}

\title{VV-NET: Voxel VAE Net with Group Convolutions for Point Cloud Segmentation}


\author{Hsien-Yu Meng\textsuperscript{1,4}, 
Lin Gao\textsuperscript{2}\thanks{Corresponding Author}, 
Yu-Kun Lai \textsuperscript{3}, 
Dinesh Manocha\textsuperscript{1}\\
	\textsuperscript{1}University of Maryland, College Park \\
	\textsuperscript{2}Beijing Key Laboratory of Mobile Computing and Pervasive Device, \\
	Institute of Computing Technology, Chinese Academy of Sciences \\
	\textsuperscript{3}School of Computer Science \& Informatics, Cardiff University\\
	\textsuperscript{4} Tsinghua University \\
    {\tt\small mengxy19@umd.edu, gaolin@ict.ac.cn, LaiY4@cardiff.ac.uk, dm@cs.umd.edu}
}

\maketitle
\ificcvfinal\thispagestyle{empty}\fi

\begin{abstract}
We present a novel algorithm for point cloud segmentation. 
\mxy{Our approach transforms unstructured point clouds into regular voxel grids, and further uses a kernel-based interpolated variational autoencoder (VAE) architecture to encode the local geometry within each voxel.
Traditionally, the voxel representation only comprises Boolean occupancy information
which fails to capture the sparsely distributed points within voxels in a compact manner. In order to handle sparse distributions of points, we further employ radial basis functions (RBF) to compute a local, continuous representation within each voxel. Our approach results in a good volumetric representation that effectively tackles noisy point cloud datasets and is more robust for learning.}
\mxy{Moreover, we further introduce group equivariant CNN to 3D, by defining the convolution operator on a symmetry group acting on $\mathbb{Z}^3$ and its isomorphic sets. This improves the expressive capacity without increasing parameters, leading to more robust segmentation results.}
%
We highlight the performance on standard benchmarks
and show that our approach outperforms state-of-the-art segmentation algorithms on the ShapeNet and S3DIS datasets.


\end{abstract}

\section{Introduction}

\noindent 3D data processing including classification and segmentation flourishes these days as 3D data can be easily captured using 3D scanners or depth cameras. It is eminent to deal with irregular and unordered data formats such as the point cloud. The processing  pipeline must also be robust towards rotation, scaling, translation and permutation on input data as mentioned in \cite{Charles_2017}. However, previous work fails to capture the internal symmetry within point clouds. We address these issues in this paper by proposing a novel representation that considers both spatial distribution of points and group symmetry in a unified framework.

In this paper, we address the problem of developing  more effective learning methods using regular data structures such as voxel-based representations, to retain and exploit spatial distributions. Typically, each voxel only contains the Boolean occupancy status (i.e. occupied or unoccupied), rather than other detailed point distributions and therefore can only capture limited details. We address this problem by investigating alternative representations, which can effectively encode the distribution of points in a voxel.


\noindent {\bf Main Results:} \mxy{We present a novel learning method for point cloud segmentation. 
The key idea is to effectively encode point distributions within each voxel. Directly treating the point distribution as a 0-1 signal is highly non-smooth, and cannot be compactly represented as per Mairhuber-Curtis theorem ~\cite{wendland_2004}.}
We instead transform an unstructured point cloud to a voxel grid. Moreover, each voxel is further subdivided into subvoxels \mxy{that interpolate sparse point samples within the voxel
by smooth Radial Basis Functions, which are symmetric around point samples as centers and positive definite.} This smooth signal can then be effectively compacted, and to achieve this we train a variational auto-encoder (VAE)~\cite{journals/corr/KingmaW13} to map the point distribution within each voxel to a compact latent space. 
Our combination of RBF and VAE provides an effective approach to representing point distributions within voxels for deep learning.

A key issue with 3D representations is to ensure that the result of point cloud segmentation does not change due to any rotations, scaling or translation with respect to an external coordinate system.
In order to capture the intrinsic symmetry of a point cloud, we use group equivariant convolutions~\cite{cohen2016group} and combine the per point feature extracted by an $mlp$ function similar to ~\cite{Charles_2017}. These group convolutions were originally proposed for 2D images and we generalize them on $\mathbb{Z}^3$ and its isomorphic sets for 3D point cloud processing. They help detect the co-occurrence in the feature space, namely the latent space of our pre-trained RBF-VAE network of voxels, and thereby improve the learning capability of our approach.   

Overall, we present VV-Net, a novel Voxel VAE network with group convolutions, and apply this to point cloud segmentation. Our approach is useful for segmenting objects into parts and 3D scenes into individual semantic objects. We have evaluated and compared its performance on standard point-cloud datasets including  \textbf{ShapeNet}~\cite{yi2016scalable} and \textbf{S3DIS}~\cite{armeni20163d}. In practice, our method outperforms the state-of-the-art methods on these datasets by $2.7\%$ and $16.12\%$ in terms of mean IoU (intersection over union), respectively.
Even when some of the ground truth data from the point cloud is labeled incorrectly, our approach is also able to compute a meaningful segmentation, as shown in Figure~\ref{fig:fail}. The novel contributions of our work include:
\begin{itemize}
    \item We develop a novel information-rich voxel-based representation for point cloud data. Point distribution within each voxel is captured using a variational auto-encoder taking RBF at the subvoxel level as input. This provides both the benefits of regular structure and capturing the detailed distribution for learning algorithms.
    
    \item We introduce group convolutions defined on the 3-dimensional data, which encode the symmetry and increase the expressive capacity of the network without increasing the number of parameters.
\end{itemize}


\section{Related Work}
\GL{There have been growing interests in 3D data processing algorithms. In this section, we give a brief overview of prior work on point cloud processing and semantic segmentation.
}
\vspace{-8mm}
\GL{\paragraph{Deep learning on 3D data.}
The point cloud is a very general representation for 3D data. A lot of pioneering research works with deep learning technologies are proposed. PointNet~\cite{Charles_2017} applies  multi-layer perceptrons to each point in the input point cloud and symmetric operations to eliminate the permutation problem. Furthermore, PointNet is robust to rotations on the input point cloud by explicitly adding a transform net to align the input point cloud. In the 3D object classification and semantic segmentation tasks, PointNet is regarded as a state-of-the-art approach. Yi et al.~\cite{yi2017learning} cluster 3D shapes by their labels in the dataset and then train a model for hierarchical segmentation.  Wang et al.~\cite{wang2018sgpn} present a similarity matrix that measures the similarity between each pair of points in the embedded space to produce the semantic segmentation map. To capture information at different scales, a commonly used approach is to capture the hierarchical information by recursive sampling or  recursively applying neural network structures~\cite{qi2017pointnet++}. In particular, the work~\cite{huang2018recurrent} applies recurrent neural networks to combine slice pooling layers, and the work~\cite{su2018splatnet} uses sparse bilateral convolutional layers as building blocks. Some methods work on 3D meshes, and strive to extract  information from graph structures generated from a mesh representation. Yu et al.~\cite{yi2017syncspeccnn} use a spectral CNN method that enables weight sharing by parameterizing kernels in the spectral domain spanned by graph Laplacian eigenbases. Verma et al.~\cite{verma2018feastnet} use graph convolutions proposed in~\cite{bruna2013spectral} to design a graph-convolution operator, which aims to establish correspondences between filter weights and graph neighborhoods with arbitrary connectivity. 
Deep learning based on variational autoencoders is also employed in \cite{Tan_2018_CVPR,Gao2019sdmnet} for mesh generation.
}
\vspace{-2mm}
\paragraph{Point cloud processing using neighborhood mining.} To address lack of connectivity, some methods use K-nearest neighbors in the Euclidean space and exploit information within local regions ~\cite{DBLP:journals/corr/DGCNN,DBLP:journals/corr/pointCNN,li2018SONet,shen2018mining}. In particular, Li et al.~\cite{li2018SONet} model the spatial distribution of point clouds by building a \textit{self-organizing map} and applying PointNet~\cite{Charles_2017} to multiple smaller point clouds. Moreover, the works~\cite{DBLP:journals/corr/DGCNN,DBLP:journals/corr/LargeScalePointCloudSemanticSegmentationWithSuperpointGraphs,li2018SONet,DBLP:journals/corr/pointCNN} use graph structures and graph Laplacian to capture the local information in the selected neighborhoods and leverage the spatial information~\cite{DBLP:journals/corr/pointCNN}.  Remil et al.~\cite{remil2017data} utilize the shape priors which are defined as point-set neighborhoods sampled from shape surfaces. However, there are many issues that make it challenging to mine the neighborhood information: First, topology information is not easy to capture with LiDAR scans, which makes it more challenging to estimate vertex normals. Second, encoding K-nearest neighborhoods in the Euclidean space may in some cases simultaneously encode two points that  do not belong to the same object (especially for the circumstance that two objects are close to each other). In our work, we do not explicitly encode the K-nearest neighborhoods in our architecture. Instead, we aim to encode the symmetry information rather than encoding the neighborhood information.

\vspace{-1mm}
\paragraph{Point cloud processing using voxels.} Some works use voxels for processing point data (e.g. ~\cite{wang2018sgpn,zhou2017voxelnet,maturana2015voxnet,qi2016volumetric}). These methods apply neural networks on voxelized data, and cannot be applied to raw point clouds directly due to their irregular and unordered data format. However, the resolution is limited by data sparsity and computational costs. For the purpose of 3D detection, Zhou and Tuzel~\cite{zhou2017voxelnet} sample a LiDAR point cloud to reduce the computation overhead and irregularity of  point distribution using farthest point sampling. In order to further reduce the imbalance of points between voxels, their method only takes into consideration densely populated voxels. It applies the point-wise feature learning function $mlp$ on each point and aggregates the features by a symmetric function. In contrast, our method does not perform sampling to eliminate the unbalanced distribution. Instead we use regular voxels along with RBF to improve the learning capabilities. 

\paragraph{Convolutions defined on groups, equivariance and transformations.}
It is known that the power of CNNs lies in the translation equivariant property, and they  exploit translational symmetries by CNN kernel weight sharing~\cite{DBLP:journals/corr/sphericalCNN}. Recently, Cohen and Wellin~\cite{cohen2016group} introduced equivariance to the combinations of $90^\circ$-rotations and dihedral flips in CNNs. They extend the theory to a steerable representation which is the composition of elementary feature types although it requires special treatment for anti-aliasing~\cite{cohen2016steerable}. Cohen et al.~\cite{DBLP:journals/corr/sphericalCNN} further introduce the spherical cross-correlation which satisfies the generalized Fourier transformation although the resulting spherical CNN requires a closed genus-0 manifold as input so that it can be projected as a spherical signal. Similarly, Weiler et al.~\cite{weiler2017learning} and 
Worrall et al.~\cite{worrall2017harmonic} design SO(2) steerable networks, although they are limited by discrete groups and are computationally expensive. All of these methods are either designed for the 2D image domain or the spherical surface domain, and none of them work directly for 3D point data. 
 
\begin{figure}[t!]
    \centering
    \includegraphics[width = \linewidth]{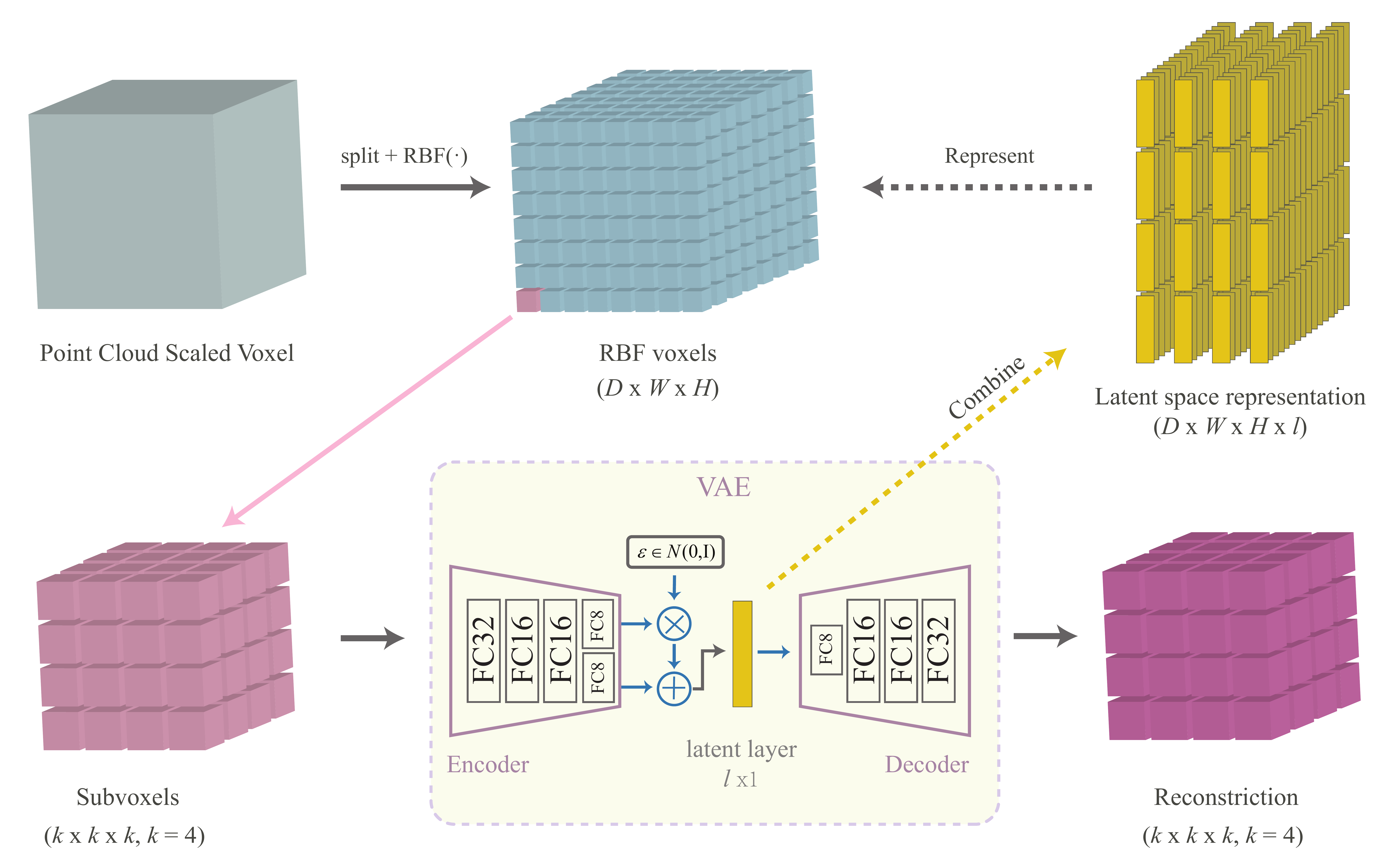}
    \caption{\textbf{Radial Basis Function interpolated Variational Auto-Encoder module.} 
    For a given point cloud, we divide it into equally spaced $D \times H \times W$ voxels, and for each voxel we further divide it into $k \times k \times k$ subvoxels, where each subvoxel value is defined by the radial basis function in Equation~\ref{eq:rbf} rather than Dirac delta function sampled by $sinc$.  The kernel of RBF is set to $\phi(||\cdot||_2^2)$ according to VAE latent distribution. For a voxel with $k \times k \times k$ subvoxels, we infer the latent space representation using a pre-trained variational auto-encoder. Finally, the point cloud can be presented as a $D \times H \times W \times l$ voxel data, where $l$ denotes the dimension of the latent space.}
    \label{fig:rbf_vae}
    \vspace{-15px}
\end{figure}

\begin{figure}[t!]
    \centering
    \includegraphics[width = \linewidth]{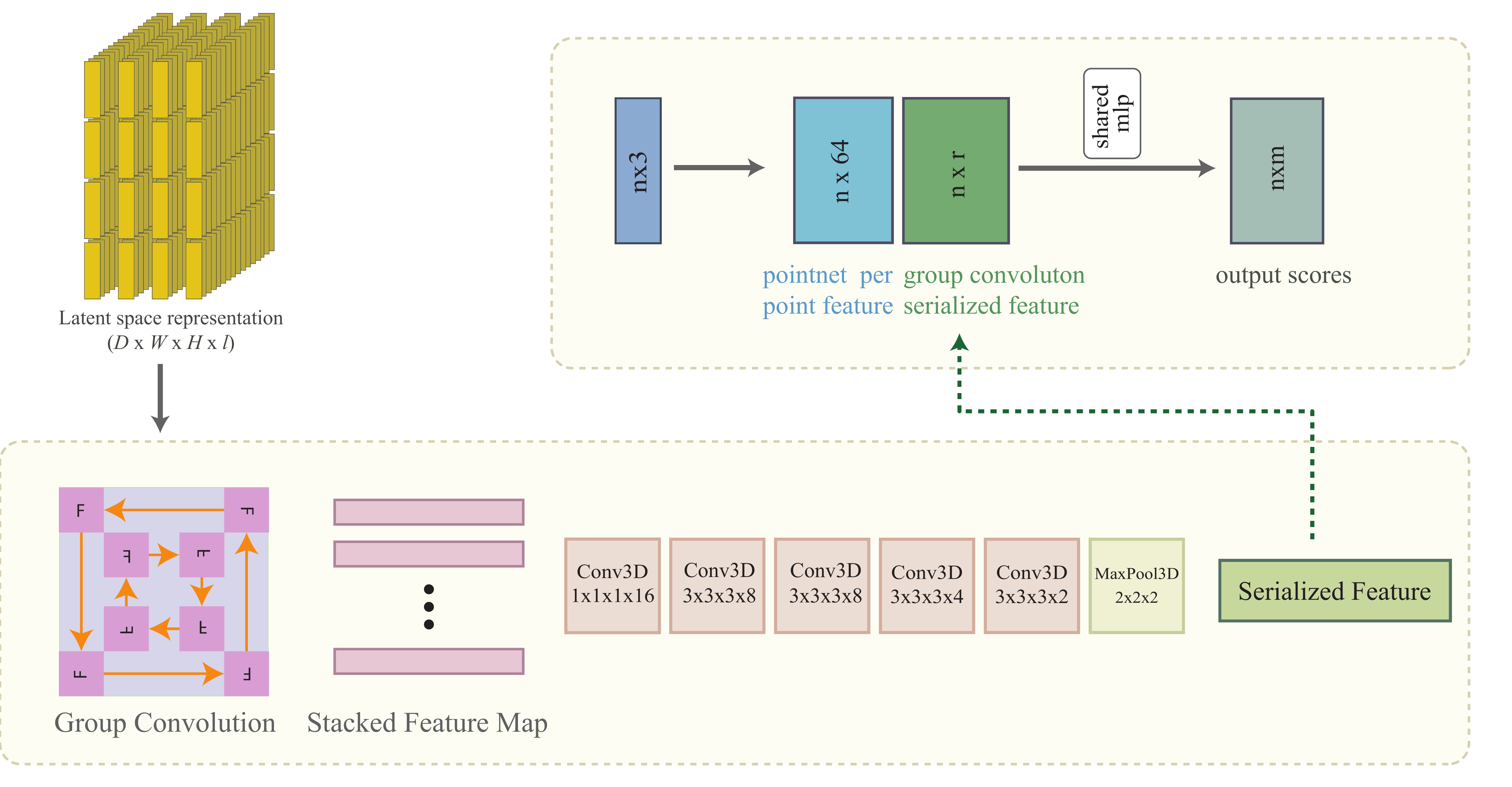}
    \caption{\textbf{Segmentation Network Architecture.} We highlight the various components of our approach. 
    The input of the network is a point cloud containing $n$ points and the latent space representation is illustrated in Figure~\ref{fig:rbf_vae}. The output is the per-class score of each point in the point cloud (for $m$ classes). We use the group convolutional module to detect the co-occurrence in the feature space (see Equation~\ref{eq:p4m}). We highlight the group $p4m$ for functions $g(m_x, m_y, m_z, r_x, r_y, r_z, t_x, t_y, t_z)$ in Equation~\ref{eq:p4m} in the bottom left figure (where $m_*$, $r_*$ and $t_*$ refer to mirroring, rotation and translation).
    A $p4m$ function has $128$ planar patches in our formulation, where each is associated with a rotation $r_x$, $r_y$, $r_z$ and mirroring $m_x$, $m_y$, $m_z$. In this figure,  we only illustrate $8$ planar patches. Each patch follows the arrow and undergoes a $90^\circ$ rotation. The patches on the outer square are mirror reflection of the patches on the inner square, and vice-versa. }
    \label{fig:network}
    \vspace{-6mm}
\end{figure}

\GL{\section{Voxel VAE Net with Group Convolutions}\label{sec:framework}}
In this section we describe the overall algorithm and highlight the various stages of the pipeline. 
\mxy{First, we illustrate the interpolation of multidimensional scattered samples, and show the intuitive motivation of VAE equipped with RBF kernel, which enjoys several advantages: symmetric and positive definite for any choice of data locations.}
Our formulation computes a better representation with an encoder-decoder scheme, instead of using the standard \{0, 1\} voxels (occupancy). Empirically, the distribution of \{0,1\} voxels is discrete and insufficient to fully capture point distributions. Moreover, its discontinuous nature makes it difficult to be learned by a deep neural network.  Second, we describe our mathematical framework based on group convolutions defined on $\mathbb{Z}^3$ and their isomorphic sets to detect the co-occurrence of features in the latent space. This increases the expressive capacity of the CNN without increasing the number of parameters and the number of layers. Third, we concatenate the $n\times64$ per-point features extracted by the $mlp$ function~\cite{Charles_2017} with the serialized features extracted by our network, where $n$ is the number of points, and $64$ is the dimension of features extracted using PointNet. Finally, after $mlp$ layers, we output the score map which indicates the probability of a point belonging to the $m$ classes as in the upper right of Figure~\ref{fig:network}, where $m$ is the number of classes in the segmentation task (e.g.~$40$ in the \textbf{ShapeNet} part segmentation task and $13$ in the \textbf{S3DIS} semantic segmentation task). 

\subsection{Symbols and Notation}
If $\mli{G}$ is a group acting on set $\mli{X}$, and $f, g : \mli{X} \rightarrow \mathbb{C} $ are actions on group $\mli{G}$, then the convolution is defined as:
\begin{equation}
    (f \ast g)(u) = \int_G f(uv^{-1})g(v)d\mu(v)
\end{equation}
where $\mu$ is Haar-measure. In this paper, we have $\mli{X} = \mathbb{Z}^{3}$, and $\mli{G}$ is the group of integer transformation, which is isomorphic to $\mathbb{Z}^{3}$. Note that this is a special case, and  $\mli{G}$ and $\mli{X}$ are usually two different sets.

In our pipeline, the input is a point cloud, represented using 3D coordinates $(x, y, z)$ in the Euclidean space. We use the symbols $(\tilde{x}, \tilde{y}, \tilde{z})$ to represent the coordinates of voxel grid.  In particular, for a given point cloud with $n$ points which encompasses 3D space with ranges $\tilde{D}$, $\tilde{H}$ and $\tilde{W}$ in the $Z$, $Y$, $X$ axes, respectively, we divide the entire point cloud into $D \times H \times W$ voxels. Therefore, the sizes of a voxel in $Z$, $Y$ and $X$ directions are: $v_D = \tilde{D}/D$, $v_H = \tilde{H}/H$ and $v_W = \tilde{W}/W$. The output of our RBF-VAE scheme is a $(D, H, W, l)$-size matrix, where $l$ represents the latent space dimension of the encoder-decoder setting. We use the notion of symmetry groups for group equivariant convolutions. Given a group $\mli{G}$, we can define a $\mli{G}$-$\mli{CNN}$ by analogy to standard $\mli{CNNs}$, by similarly defining the function $\mli{G}$-$\mli{convolution}$ on the group $\mli{G}$.

\subsection{RBF-VAE Scheme}\label{sec:rbf-vae-sheme}
\mxy{The traditional voxel representation can be deemed as a 0-1 signal $f$ sampled at each grid point with spacing $v_D, v_H, v_W$ along each dimension by Whittaker - Shannon interpolation formula. }
\mxy{Applying Fourier transformation to such signal $f$ involving a combination of Dirac delta functions produces a dense distribution in the frequency domain, forming a Haar-space (Chebyshev space),  which cannot be effectively compacted, according to Mairhuber-Curtis theorem \cite{wendland_2004}. }
Instead of Boolean occupancy information, we evaluate grid value at $p$ as a linear combination of radial basis functions:
\begin{equation}\label{eq:rbf_sum}
    f(p) = \sum_{j = 1}^{N}w_j{\phi(||p - v_j||_2^2)}
\end{equation}
where $N$ is the number of data points, $w_j$ is a scalar value and $\phi(\cdot)$ is a symmetry function about each data point and is positive definite according to \textit{Bochner theorem}.
\mxy{We measure the point distribution over $k \times k \times k$ subvoxels by using a variational auto-encoder, leading to an $l$-dimensional latent space for each voxel, which is not only compact but also captures the spatial distribution of points. }
Overall, the voxel representation size for the entire point cloud is $D \times H \times W \times l$, which is  more detailed than the standard $D \times H \times W$ volumetric representation.  

\begin{table*}
\caption{\textbf{ShapeNet experiment settings to test the performance of each module:} Our VAE module is illustrated in Figure~\ref{fig:rbf_vae}, and the group convolutional module is highlighted in Figure~\ref{fig:network}. 
We present the parameters used for our approach (group-conv + RBF-VAE) and with one module disabled, namely only RBF-VAE without group-conv and group-conv with \{0, 1\} voxels. The input subvoxel (for VAE-based) or voxel (for non-VAE based) resolutions are fixed to $64\times64\times 64$.}

\begin{adjustbox}{max width=\textwidth}
    \begin{tabular}{llll}
    \hline
        Experiment      & input of VAE                        & output of VAE                        & input of group conv              \\
        \hline
        
        
        RBF-VAE         & $64\times64\times64$ RBF voxel      & $16\times16\times16\times8$ latent voxel&  None \\
        
        group-conv + \{0,1\} voxel & None   & None & $64 \times 64 \times 64 $ \{0,1\} voxel\\
        (\textbf{Our})group-conv + RBF-VAE  &  $64\times64\times64$ RBF voxel & $16\times16\times16\times8$ latent voxel &  
                               $16\times 16\times 16\times 8$ latent voxel \\      
        \hline
    \end{tabular}
    \label{tab:exp}
    \end{adjustbox}\vspace{-3mm}
\end{table*}

\begin{table*}
\caption{\textbf{Results on ShapeNet part segmentation:} We highlight the instance average mIoU and mIoU scores for all the categories on point cloud labeling using prior algorithms and our method. Note that the comparison performances listed below are reported by PointNet~\cite{Charles_2017}, RSN~\cite{huang2018recurrent}, SO-Net~\cite{li2018SONet}, SynSpecCNN~\cite{yi2017syncspeccnn} and SPLATNET~\cite{su2018splatnet}, respectively. The numbers in bold show the best performances for different object categories. Furthermore, in our experiments, we highlight the results that outperform the state-of-the-art method. The $3$ experiments listed in the bottom correspond to the experiment settings in Table~\ref{tab:exp}. Overall, with both the RBF-VAE module and the group convolutional module, our method outperforms the state-of-the-art method by $2.5\%$ in terms of mean IoU. If we replace the RBF-VAE module with the standard \{0,1\} voxel VAE module, the training does not converge because the point data is too sparse.
Moreover, if we remove the group convolutional module or the RBF-VAE module
from our complete pipeline,  mIoU would drop by $1.3\%$ or $1.4\%$ respectively. Note that the \textit{Motor} and \textit{Car} categories are challenging as they each contain 4 or more parts. Nonetheless, our method shows significantly better performances.
}
\begin{adjustbox}{max width=\textwidth}

\begin{tabular}{llllllllllllllllll}
\hline

                              & Mean IoU      & Aero          & Bag           & Cap  & Car           & Chair & Ear           & Guitar        & Knife         & Lamp & Laptop & Motor         & Mug           & Pistol        & Rocket        & Skate & Table \\ 
\hline

PointNet~\cite{Charles_2017}
                                        & 83.7          & 83.4          & 78.7          & {82.5}        & 74.9          &
                                         {89.6}         & 73.0          & 91.5          & 85.9          & {80.8}        & 
                                         {95.3}         & 65.2          & 93.0          & 81.2          & 57.9          & 
                                         72.8           & 80.6  \\

RSN~\cite{huang2018recurrent}
                                        & 84.9          & 82.7          & 86.4          & 84.1          & 78.2          &
                                          {90.4}        & 69.3          & 91.4          & 87.0          & {83.5}        &
                                          95.4          & 66.0          & 92.6          & 81.8          & 56.1          &
                                          75.8          & 82.2\\

SO-Net~\cite{li2018SONet}
                                       & 84.6          & 81.9          & 83.5          & 84.8          & 78.1          &
                                         \textbf{90.8} & 72.2          & 90.1          & 83.6          & 82.3          &
                                         95.2          & 69.3          & 94.2          & 80.0          & 51.6          &
                                         72.1          & 82.6 \\

SyncSpecCNN~\cite{yi2017syncspeccnn} & 84.74          & 81.55         & 81.74         & 81.94           & 75.16          & 90.24 &
                                               74.88          & \textbf{92.97}& 86.10         & \textbf{84.65}  & {95.61}        & 66.66 &
                                               92.73          & 81.61         & 60.61         & 82.86           & 82.13\\

SPLATNET-3D~\cite{su2018splatnet}      
                                            & 84.6           & 81.9          & 83.9          & \textbf{88.6}& 79.5           & 90.1 
                                            & 73.5           & 91.3          & 84.7          & 84.5         & \textbf{96.3}  & 69.7 
                                            & 95.0           & 81.7          & 59.2          & 70.4         & 81.3 \\
                                            
\hline
RBF-VAE                                     & \textbf{86.1}  & 82.3          & \textbf{86.6} & 82.4        & \textbf{81.7}  & 
                                              87.7           & \textbf{77.1} & 91.2          & 83.7        & 77.5           & 94.0  &
                                              \textbf{71.0}  & \textbf{96.1} & \textbf{86.6} & 56.1        & \textbf{87.8}  & \textbf{89.5}\\
                                              



group-conv + \{0,1\} voxel &
                           \textbf{86.0} & 82.1 & 68.9 & 83.8 & \textbf{80.9} & 87.8 & \textbf{81.2} & 91.2 &
                            78.4 & 77.4 & 94.5 & \textbf{72.8} & \textbf{98.0} & \textbf{86.0} & 53.8 & \textbf{83.9} & \textbf{90.0}\\
group-conv + RBF-VAE                                 
                                        & \textbf{87.4} & \textbf{84.2} & \textbf{90.2} & 72.4          & \textbf{83.9} & 
                                          88.7          & \textbf{75.7} & {92.6}        & \textbf{87.2} & 79.8 & 94.9   & 
                                          \textbf{73.4} & {94.4}        & \textbf{86.4} & \textbf{65.2} & \textbf{87.2} & \textbf{90.4} \\ 

\hline
          
\end{tabular}
\label{tab:shapenet}
\end{adjustbox}\vspace{-3mm}
\end{table*}

\subsubsection{Radial Basis Functions}
To map discrete points to a continuous distribution, 
we use radial basis functions to estimate their contributions within each subvoxel:
\begin{equation}\label{eq:rbf}
    f(p) = \max_{v\in V}\bigg(\exp{\frac{-||p - v||_2^2}{2 \sigma^2}}\bigg).
\end{equation}
Here $V$ represents the set of points, $p$ is the center of the subvoxel, and $\sigma$ is a pre-defined parameter, usually is a multiple of the subvoxel size. In principle all the points in $V$ may affect the value of $f(p)$, it is the point closest to $p$ that is dominant. As a result $f(p)$ can be evaluated efficiently.
The formulation here is based on the commonly used Gaussian RBF kernel. Empirically, the kernel used in RBF, i.e. $\phi(||\cdot||_2^2)$ has the same form as VAE latent variable distribution. Furthermore, we show the comparison results of different kernels in Section ~ \ref{sec:rbf}.

\subsubsection{Variational Auto-Encoder}
Our approach uses the approach highlighted in~\cite{journals/corr/KingmaW13} to model the probabilistic encoder and the probabilistic decoder. The encoder aims to map the posterior distribution from datapoint $X_{(D_i, H_i, W_i)}$ to the latent vector $Z_{(D_i, H_i, W_i)}$, where $(D_i, H_i, W_i)$ represents $k \times k \times k$ subvoxels and is denoted as $K_i$. And the decoder produces a plausible corresponding datapoint $\hat{X}_{K_i}$ from a latent vector $Z_{K_i}$. In our setting, the datapoint $X_{K_i}$ is represented by RBF kernel subvoxels as formulated in Equation \ref{eq:rbf}. The total loss function of our model can be evaluated as : 
\small
\begin{equation}
    \begin{split}
    \mli{Loss} & = \sum_{K_i \in (D, H, W)}  {E_{Z_{K_i}}[\log{P(X_{K_i}^{(i)} | Z_{K_i})}]}  \\
    & - D_{KL}{(q_\phi(Z_{K_i}|X_{K_i}^{(i)}) || P_\theta(Z_{K_i}))}  \\
    & + D_{KL}(q_\phi(Z_{K_i}|X_{K_i}^{(i)}) || P_\theta(Z_{K_i}|X_{K_i}^{(i)})) \\ 
    \end{split}
\end{equation}
\normalsize
where we sample $Z_{K_i}|X_{K_i}$ from $Z_{K_i}|X_{K_i} \sim \mathcal{N}(\mu_{Z_{K_i}|X_{K_i}}, \Sigma_{Z_{K_i}|X_{K_i}})$ and sample $X_{K_i}|Z_{K_i}$ from $X_{K_i}|Z_{K_i} \sim \mathcal{N}(\mu_{X_{K_i}|Z_{K_i}}$, $ \Sigma_{X_{K_i}|Z_{K_i}})$, $q_\phi(Z_{K_i}|X_{K_i})$ indicates the encoder network and $P_\theta(X_{K_i}|Z_{K_i})$ indicates the decoder network. Note that the latent variable $Z_{K_i}$ only captures the spatial information within a single voxel by the variational auto-encoder scheme. For a pre-trained VAE module, we infer each voxel from the fixed-parameter VAE and compute the final point cloud representation of size $D \times H \times W \times l$, where $l$ is the latent space size of the pre-trained VAE module. 
\GL{The variational auto-encoder captures point data distribution within a voxel in a more compact manner. 
This not only reduces memory footprint, but also makes our learning algorithm more efficient. The VAE has significantly better generalizability than AE due to the prior distribution assumption, and avoids potential overfitting to the training set.}

\begin{table*}
\caption{\textbf{Results of Semantic Segmentation on the S3DIS Dataset.} Our underlying metric is Intersection over Union (IoU) calculated on the points, evaluated on the benchmark~\cite{armeni20163d}. One metric is different between Table~\ref{tab:s3dis} and Table~\ref{tab:S3DIS_AP0.5}, IoU in  Table~\ref{tab:s3dis} and $AP_{0.5}$ in Table~\ref{tab:S3DIS_AP0.5}, following the practice of existing papers. We report both metrics while most previous works choose to report one or the other. The numbers in bold face fonts indicate the best performances and we highlight the numbers in our experiments if the results outperform the state-of-the-art methods. 
Notice that the full pipeline (last experiment) outperforms only using RBF-VAE by $1.8\%$ and only using group-conv by $6.33\%$.
Note that the performances of PointNet~\cite{Charles_2017}, Engelmann~\cite{engelmann2017exploring} and SPG~\cite{DBLP:journals/corr/LargeScalePointCloudSemanticSegmentationWithSuperpointGraphs}  are reported in~\cite{DBLP:journals/corr/LargeScalePointCloudSemanticSegmentationWithSuperpointGraphs}. The RSN~\cite{huang2018recurrent} performance is reported in their paper. 
}

\begin{adjustbox}{max width = \textwidth}
\begin{tabular}{llllllllllllllll}
\hline
                & overall ACC    & Mean IoU       & ceiling        & floor         & wall          & beam           & column         & window         & door           & chair         & table          & bookcase & sofa          & board          & clutter        \\
\hline
PointNet~(\cite{Charles_2017})     
                & 78.5           & 47.6           & 88.0           & 88.7          & 69.3          & 42.4           & 23.1           & 47.5           & 51.6           & 42.0          & 54.1           & 38.2          & 9.6           & 29.4           & 35.2           \\

Engelmann~(\cite{engelmann2017exploring})
                & 81.1           & 49.7           & 90.3           & 92.1           & 67.9         & 44.7           & 24.2
                & 52.3           & 51.2           & 47.4           & 58.1           & 39.0         & 6.9            & 30.0
                & 41.9\\
                
SPG~(\cite{DBLP:journals/corr/LargeScalePointCloudSemanticSegmentationWithSuperpointGraphs})
                & 85.5           & 62.1           & 89.9           & {95.1}         & {76.4}       & 62.8           & 47.1           & 55.3           & 68.4         & \textbf{73.5}  & 69.2           & \textbf{63.2} & \textbf{45.9} & 8.7            & 52.9           \\

RSN~(\cite{huang2018recurrent})
                & 59.42              & 51.93         & \textbf{93.34} & \textbf{98.36} & \textbf{79.18} & 0.00          & 15.75
                & 45.37          & 50.10         & 65.52          & 67.87          & 22.45          & 52.45         & 41.02
                & 43.64\\

\hline



RBF-VAE         & \textbf{85.98} & \textbf{75.40} & 85.01         & 95.52          & 71.58         & \textbf{73.81} & \textbf{60.91} 
                & \textbf{61.54} & \textbf{74.38} & 65.67         & 67.59          & 61.47         & 26.11          &  38.72       
                & \textbf{56.16}\\

group-conv + \{0,1\} voxel
                & 81.45 & \textbf{68.70} & 83.27 & 93.95 & 59.37 & \textbf{64.35} & 40.23 & 54.06 & 66.48 & 
                65.20 & 63.52 & 41.48 & 20.37 & 16.21 & 47.41\\
group-conv + RBF-VAE         
                & \textbf{87.78} & \textbf{78.22} & 87.64         & 95.36          & 74.80         & \textbf{75.04} & \textbf{68.03}
                & \textbf{71.33} & \textbf{76.87} & 72.67         & \textbf{70.08} & 61.97         & 33.56          & \textbf{49.81}
                & \textbf{60.00}\\
\hline
\end{tabular}
\label{tab:s3dis}
\end{adjustbox}
\end{table*}

\begin{table*}
    \caption{\textbf{Results of Semantic Segmentation on the S3DIS Dataset with $AP_{0.5}$.} The metric is average precision (AP(\%)) with IoU threshold 0.5. Note that the complete pipeline (group-conv+RBF-VAE) achieves the best performance, outperforming both state-of-the-art work and with one of our modules disabled.
    The result of Armeni~\cite{armeni20163d} is for 3D object detection and IoU is calculated on 3D bounding boxes, while SGPN and ours are based on point cloud datasets. Note that the comparison performances listed below are reported in SGPN~\cite{wang2018sgpn}. }
\begin{adjustbox}{max width = \textwidth}
\begin{tabular}{llllllllllllll}
\hline
                 & Mean IoU($AP0.5$)       & ceiling        & floor         & wall          & beam           & column         & window         & door           & chair         & table          & bookcase & sofa          & board        \\
\hline
Armeni~(\cite{armeni20163d}) 
                         & 49.93   & 71.61   & 88.70   & 72.86          & 66.67          & \textbf{91.77} &
                            25.92   & 54.11   & 16.15   & 46.02          & 54.71          & 6.78           & 
                            3.91\\
SGPN~(\cite{wang2018sgpn})   
                          & 54.35           & 79.44             & 66.29            & \textbf{88.77} & 77.98          & 60.71          &
                            66.62   & 56.75   & 40.77   & 46.90          & 47.61          & 6.38           &
                            11.05\\
                            
\hline

RBF-VAE                   & \textbf{79.00}   & \textbf{88.73} & \textbf{97.43}  & 77.20          & \textbf{79.91} & 67.27
                          &  62.39           & \textbf{81.36} & \textbf{67.08}  & \textbf{74.68} & \textbf{55.38} & \textbf{37.08}
                          & \textbf{33.05} \\

group-conv + \{0,1\} voxel  
& \textbf{72.66} & \textbf{88.98} & \textbf{95.32} 
& 64.13 & 67.74 & 49.21 & 55.35 & \textbf{74.02} &
\textbf{64.34} & \textbf{68.38} & 29.11 & \textbf{22.58} & \textbf{13.33} \\

group-conv + RBF-VAE        & \textbf{82.17}  & \textbf{91.68}  & \textbf{96.54}  & 80.38          & \textbf{80.68} & 71.87
                          & \textbf{72.94}  & \textbf{85.81}  & \textbf{73.86}  & \textbf{76.76} & \textbf{57.68} & \textbf{43.82}
                          & \textbf{46.35} \\

\hline
    \end{tabular}
    \label{tab:S3DIS_AP0.5}
    \end{adjustbox}
\end{table*}

\subsection{Symmetry Group and Equivariant Representations}
In this section, we present our algorithm to compute the equivariant representations using the symmetry groups. The goal is to build on the VAE based voxel representation and 
detect the co-occurrence in features with filters in the CNN. The ultimate goal is to enhance the network expressive capacity without increasing the number of layers or the filter sizes in the standard CNN.
The work~\cite{cohen2016group} illustrates these issues in the current generation of neural networks, where the representation spaces  have minimal internal structure. To address this issue, we use symmetry groups and equivariance CNN to perform efficient data processing. In this case, \textit{G-CNN} is defined in the linear $G$-space, where each vector in the $G$-space has a \textit{pose}, and can be transformed by an element from a group of transformations \textit{G}. Particularly, \textit{G-convolution} corresponds to an operation that helps a filter in \textit{G-CNN} to detect the co-occurrence in features. The transformation in \textit{G-space} is structure preserving.
We extend the formulation of  $G$-space presented in~\cite{cohen2016group} which was defined for 2D images to 3D. In particular, we define and use $p4$ and $p4m$ as symmetry groups on $\mathbb{Z}^3$. Furthermore, we show the group equivariant convolution on $\mathbb{Z}^3$ and the underlying CNN is a function on the group. When we apply $90^\circ$  rotations on a function on $p4$, the simplified results of this operation are shown in Figure \ref{fig:network}.

\subsubsection{Group $p4$}
The group $p4$ is comprised of all compositions of translations and rotations by $90^\circ$ about any center of rotation in a 3D grid. We can parameterize the group $p4$ in terms of $r_x, r_y, r_z,  t_x, t_y, t_z$ where $r_*$ and $t_*$ are rotations and translations w.r.t. axis *, respectively. Here $*$ refers to either $X, Y$ or $Z$. This can be formulated as $g(r_x, r_y, r_z, T) = R_x \times R_y \times R_z \times T$, where $R_*$ is the rotation matrix which rotates around the axis * by $\frac{\pi \cdot r_*}{2}$, and $T$ is the translation matrix which translates along the $X$, $Y$, $Z$  axes by $t_x$, $t_y$, $t_z$, respectively. Here, $0\leq r_x \leq 4$ , $0\leq r_y\leq 4$, $0\leq r_z \leq 4$ 
and $(t_x, t_y, t_z) \in \mathbb{Z}^3$. The group operation is performed using matrix multiplication. As mentioned above, the composition of two functions and the inverse function can be easily formulated in terms of $(r_x, r_y, r_z, t_x, t_y, t_z)$, hence the operation defines a symmetry group. The group $p4$ acts on points in $\mathbb{Z}^3$ (voxel coordinates) by multiplying the matrix $g(r_x, r_y, r_z, t_x, t_y, t_z)$ by homogeneous coordinates of a point.

\subsubsection{Group $p4m$} 
Here, we extend the group $p4$ and construct a symmetry group $p4m$ defined on $\mathbb{Z}^3$, which also includes mirroring (reflection) along axis aligned planes. More formally, we have the following lemma:

\begin{lemma}
The group $p4m$ is comprised of all compositions of transformations,  rotations by $90^\circ$ about any center of rotation in the grid, and mirror reflections (i.e. $p4$ plus mirroring). As the group $p4$ formulated above, we can parameterize the group $p4m$ in terms of integers $(m_x, m_y, m_z, r_x, r_y, r_z, t_x, t_y, t_z)$ as $R_{mx} \times R_{my} \times R_{mz} \times T$, where $R_{mx}$ is formulated as below:
\begin{equation}\label{eq:p4m}
        R_{mx} = \begin{bmatrix}
            (-1)^{m_x}\cos{(r_x \frac{\pi}{2})} & -(-1)^{m_x}\sin{r_x \frac{\pi}{2}} & 0 & 0\\
            \sin{(r_x\frac{\pi}{2})} & \cos{(r_x\frac{\pi}{2})} & 0 & 0\\
            0 & 0 & 1 & 0\\
            0 & 0 & 0 & 1\\        
        \end{bmatrix}, \\
\end{equation}
and $m_*$ indicates mirroring, $m_x \in \{0, 1\}$, $m_y \in \{0, 1\}$, $m_z \in \{0, 1\}$, $0\leq r_x \leq 4$, $0\leq r_y\leq 4$, $0\leq r_z \leq 4$  and $(t_x, t_y, t_z) \in \mathbb{Z}^3$.
The group $p4m$ is a symmetry group.
\end{lemma}
As illustrated in the bottom left of Figure~\ref{fig:network}, there are $128$ 3D patches 
that undergo rotation and translation transformations. The rich transformation structure arises from the group operation $p4m$. Our group operation holds the property of a symmetry group.
For implementation, the group convolution with $90^\circ$- rotations is employed by copying the transformed filters with different rotation-flip combinations ($R_{mx} \times R_{my} \times R_{mz}$). For $R_{mx}$ we have $4 \times 2$ combinations (4 choices for $90^\circ$ rotation, and whether reflection is applied, along the $X$ axis). As illustrated in Figure~\ref{fig:network}, patches are stacked to form a 5D tensor $(B \times (D - K_D)  \times (H - K_H) \times (W - K_W) \times (P \cdot C))$, where $B$ represents batch size, $D$, $H$, $W$ are the voxel sizes along $X$, $Y$, $Z$ axes mentioned in Section~\ref{sec:rbf-vae-sheme}. $K = (K_D, K_H, K_W, K_{C_{in}}, K_{C_{out}})$ is the kernel size used in the 3D CNN , $P$ is the total patch number, and $C_{in}$ and $C_{out}$ are the numbers of input and output channels of the 3D CNN.  We also developed an efficient approach to reducing the memory footprint, where 3D rotation-flip combinations are constructed based on applying 2D rotation-flip operations along arbitrary axes.  



\section{Implementation and Performance}\label{sec:exp}
Our network architecture is shown in Figure \ref{fig:network}. 
\GL{The RBF-VAE module and the segmentation module are trained separately and the RBF-VAE module is trained firstly. There are two reasons of training separately: the loss of RBF-VAE module is voxel-wise and thereby the whole network could benefit little from it; the memory consumption is saved to $1/8$ of its joint training size to make it perform on a typical Nvidia 1080Ti GPU. Our network is trained with 100 epochs and batch size 24. The inference time of our network is about 210 ms per frame on the S3DIS dataset.}
We have evaluated our segmentation method VV-Net on two datasets: \textbf{ShapeNet}~\cite{yi2016scalable} and \textbf{S3DIS}~\cite{armeni20163d}, respectively. Moreover, we demonstrate the effectiveness of each module used in our approach. First, we highlight the performance difference by alternating between the standard \{0,1\} voxel VAE module and our novel RBF voxel VAE module. Second, we evaluate the expressive capacity of the group convolutional network module. All these results and comparisons are highlighted in
Table~\ref{tab:shapenet}, Table~\ref{tab:s3dis} and Table~\ref{tab:S3DIS_AP0.5} with the parameter settings of our approach for part segmentation given in Table~\ref{tab:exp}. Our code repository is released on \href{https://github.com/xianyuMeng/VV-Net-Voxel-VAE-Net-with-Group-Convolutions-for-Point-Cloud-Segmentation}{Github}.

\subsection{Part segmentation}
\begin{figure}
    \centering
    \includegraphics[width = 0.8\linewidth]{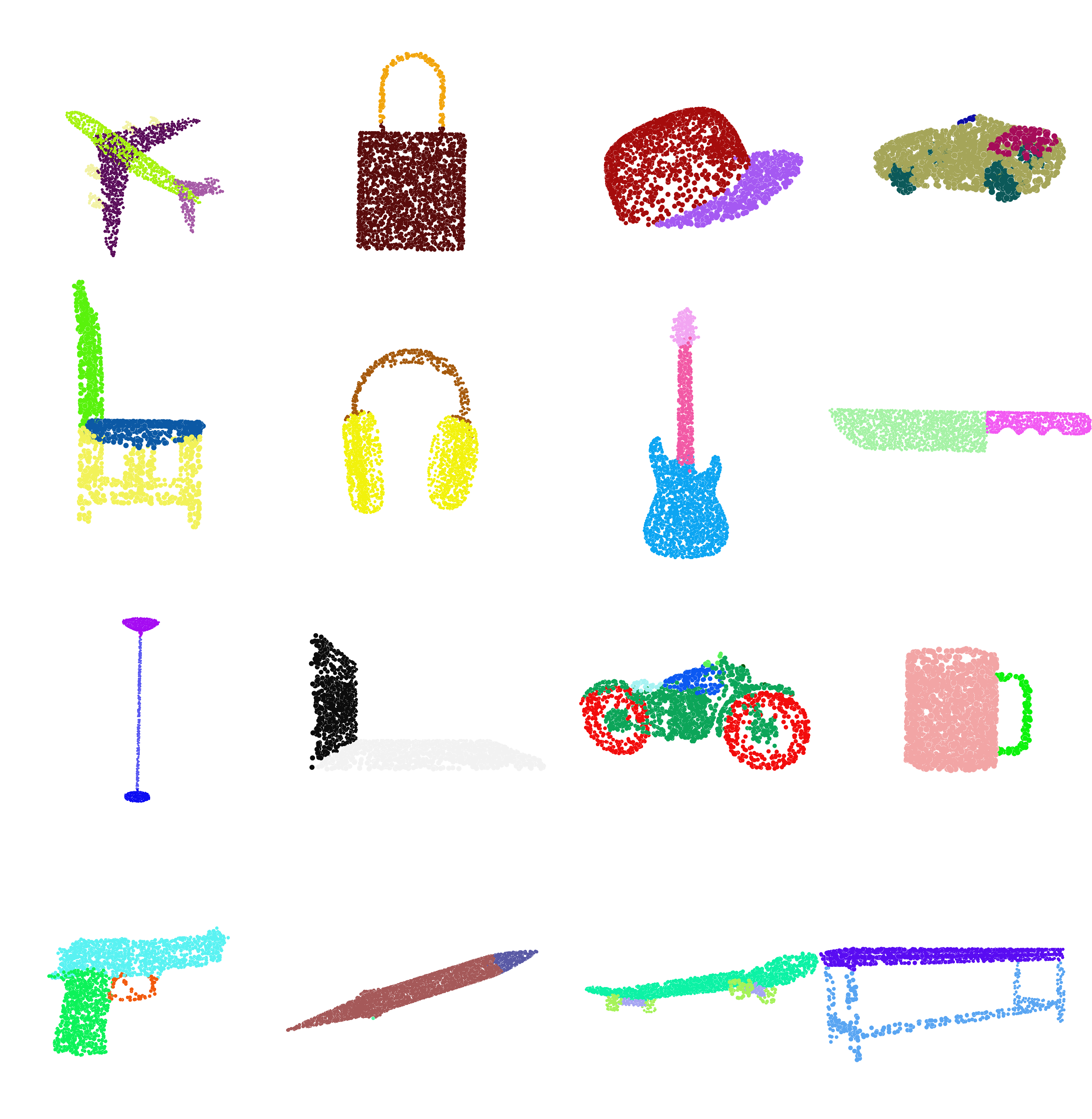}
    \caption{\textbf{Part Segmentation Results on ShapeNet.} Note that \textit{Car} and \textit{Motor} have lower performance  than most other categories in Table~\ref{tab:shapenet}. This is partly because there are more parts in these categories: $4$ labels for \textit{Car} and $5$ labels for \textit{Motor}.
    }
    \label{fig:success}
    \vspace{-7px}
\end{figure}

\begin{figure}
    \centering
    \includegraphics[width = \linewidth]{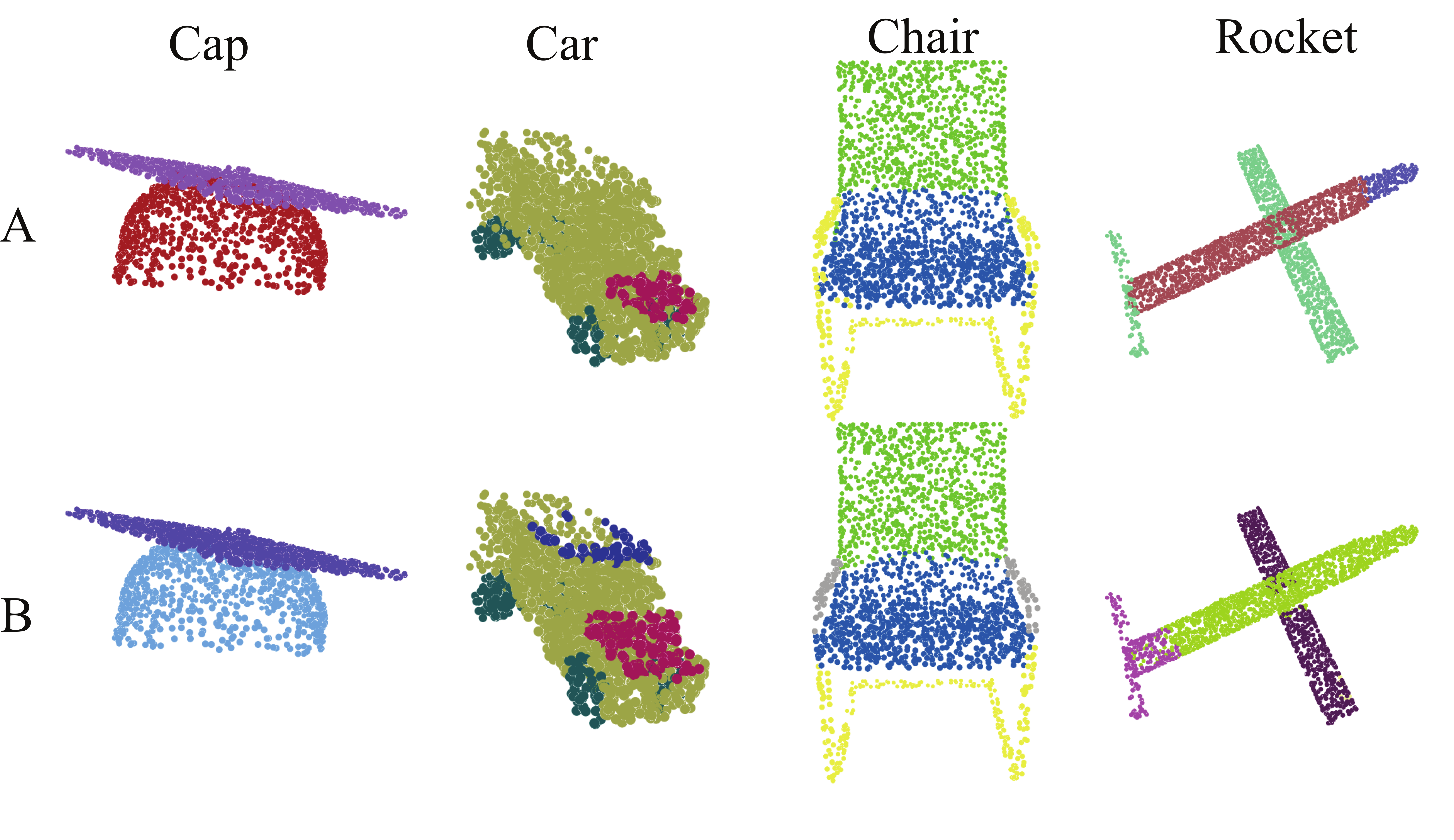}
    \caption{\textbf{{Failure} Cases of  Our Algorithm on ShapeNet Part Segmentation.} 
    The top row shows the ground truth and the bottom row is our segmentation results. Our network predicts the \textit{Cap} to be a \textit{Table}, where dark blue indicates the table top and the light blue indicates the table legs. In the second column, the dark blue indicates the top of the car. In the third column, our network segments the chair armrest while the ground truth does not. In the last column, our network predicts the \textit{Rocket} to be an \textit{Airplane}. Notice that in the last column, even a person would find it difficult to distinguish the \textit{Rocket} from the \textit{Airplane}.
    }
    \label{fig:fail}
    \vspace{-15px}
\end{figure}

Part segmentation is a challenging 3D analysis task, which aims to segment a given 3D scan into meaningful segments. We evaluate our algorithm and highlight the performance in Table~\ref{tab:shapenet} on a large-scale \textbf{ShapeNet} dataset, which contains $16,881$ shapes from $16$ categories, and annotated with $50$ parts in total. Some examples of the results of our approach are shown in Figure~\ref{fig:success}. Figure~\ref{fig:fail}(top) demonstrates the ground truth of the dataset, and we can notice that each category is labeled with two to five parts. As described in \cite{Charles_2017}, we also formulate our problem as per-point multi-label classification. The loss function is cross entropy function defined as below:
\begin{equation}
    \mli{Loss} = -\Sigma_{l}^{L}{g_l\log{p_l}},
\end{equation}
where $L$ is the number of labels, $g$ is the probability of ground truth label and $p$ is the probability of each label.
The evaluation metric is mIoU (mean IoU)  on points, following the formula in \cite{Charles_2017}: if the union of groundtruth and prediction points is empty, then we count the corresponding label IoU as 1, since we have $50$ parts and $16$ shape categories, we compute the category IoU as the average instance IoU on the category.

In our experiment, $(D, H, W) = (16, 16, 16)$, $k = 4$ , $\sigma = \min(v_W, v_H, v_D)$ and $l = 8$, where we capture the $4 \times 4 \times 4$ subvoxels with $8$ latent variables inferred from the variational auto-encoder.
We highlight the performance of various combinations of different modules. The results corresponding to \textit{group-conv + RBF-VAE} highlights VV-Net's performance based on combining  RBF kernel with VAE scheme and the group convolutional neural network module. This version of our algorithm  outperforms state-of-the-art 
RSN~\cite{huang2018recurrent} by $2.5\%$ (mIoU) and it is better than RSN in $12$ out of $16$ categories.

In order to demonstrate the benefits of individual components, we perform an ablation study. For fair comparison, the same $64\times64\times 64$ resolution of subvoxels (for VAE-based) or voxels (for non-VAE based) is used.  The implementation of \textit{group-conv+RBF-VAE} (our method) outperforms only using \textit{RBF-VAE} by $1.3\%$ (mIoU) and is better in $11$ out of $16$ categories. Our method also outperforms only using \textit{group-conv} by $1.4\%$ (mIoU) and is better in $13$ out of $16$ categories. We also compare RBF-VAE with VAE on the \{0, 1\} occupancy grid. Since the point data is sparse, training on the \{0, 1\} VAE does not converge. This shows the necessity and benefits of RBF-VAE. 

\vspace{-3px}
\subsection{Semantic segmentation of scenes}
We also evaluate the performance on Stanford 3D semantic parsing dataset \cite{armeni20163d}, which consists of $6$ types of benchmarks.  Each point in the data scan is annotated with one of the semantic labels from $13$ categories.
In our experiment, $(D, H, W) = (16, 16, 32)$, $k = 4$,  $\sigma = 5 \cdot \min(v_W, v_H, v_D)$ and $l = 8$. Table~\ref{tab:s3dis} highlights the results (category IoU, overall accuracy and mean IoU) of semantic segmentation on the \textbf{S3DIS} dataset. Furthermore, Table \ref{tab:S3DIS_AP0.5} indicates the results of AP (average precision) metric with IoU threshold $0.5$. Our implementation of \textit{group-conv + RBF-VAE} outperforms state-of-the-art SPG~\cite{DBLP:journals/corr/LargeScalePointCloudSemanticSegmentationWithSuperpointGraphs} by $16.12\%$ of \textit{Mean IoU} metric.
Our method (\textit{group-conv + RBF-VAE}) also achieves better performance than either only using \textit{group-conv} or only using \textit{RBF-VAE}, as reported in the bottom rows of Tables~\ref{tab:s3dis} and \ref{tab:S3DIS_AP0.5}. 
Table~\ref{tab:s3dis-iou} compares our method with methods reporting mean IoU and also shows the superior performance of our method.

\begin{table}
    \centering
    \begin{tabular}{lll}
        \hline
         & Mean IoU\\
         \hline
        PointCNN~\cite{DBLP:journals/corr/pointCNN} & 62.74 \\
        PointSIFT~\cite{jiang2018pointsift}         & 70.23 \\
        Ours                                                & 78.22\\
        \hline
    \end{tabular}
    \caption{\textbf{Results on Semantic Segmentation in S3DIS Dataset.} We compare the results  with~\cite{DBLP:journals/corr/pointCNN} and~\cite{jiang2018pointsift} using the mean IoU metric. }
    \label{tab:s3dis-iou}
    \vspace{-10px}
\end{table}

\subsection{Robustness test}
We have also evaluated the performance and robustness of our method by removing some of the points in the original data.
In particular, we sample the \textbf{ShapeNet} by farthest point sampling and use different missing data ratios. We evaluate the performance and accuracy of the resulting datasets. Table \ref{tab:robust} shows the result of our robustness test. This indicates that our approach is not sensitive to missing samples.

\begin{table}[]
    \centering
    \begin{tabular}{lll}
    \hline
        Missing Data Ratio & Accuracy  \\
        \hline
        0\%      & 92.47\\
        75\%     & 92.48\\
        87.5\%   & 91.70\\
        \hline
    \end{tabular}
    \caption{\textbf{Robustness Test on ShapeNet Part Segmentation Task.} In this evaluation, the point clouds are sampled by farthest point sampling. We test the robustness of our VV-Net network towards missing points. We report the mean accuracy for different missing data ratios. Our approach only has $0.77\%$ accuracy loss, even missing $87.5\%$ of the point cloud data. }
    \label{tab:robust}
    \vspace{-10px}
\end{table}

\subsection{Comparison of Different RBF Kernels}\label{sec:rbf}

Our RBF function is used to map the distance to each point to its  influence. We compare the Gaussian kernel in our method with the inverse quadratic function kernel. With this kernel, the subvoxel function value at position $p$ is defined as:
\begin{equation}
    f(p) = \max_{v\in V}\bigg(\frac{1}{1 + \sigma^2 \cdot ||p - v||_2^2}\bigg).
\end{equation}
Here $V$ represents the set of points, $p$ is the center of the subvoxel, and $\sigma$ is a pre-defined parameter, usually a multiple of the subvoxel size. The results are shown in Table~\ref{tab:rbf_test} where using Gaussian kernel achieves better performance.

\begin{table}[]
\centering
    \begin{tabular}{lll}
     & Overall Acc & mean IoU\\
     \hline
     Gaussian & 87.78 & 78.22 \\ 
     inverse quadratic & 78.82& 65.04 \\
     \end{tabular}
     \caption{\textbf{RBF Kernel Function Comparison on S3DIS Semantic Segmentation Task.} We compare the Gaussian kernel with the inverse quadratic function.}
     \label{tab:rbf_test}
     \vspace{-10px}
\end{table}

\subsection{Ablation study}
Table~\ref{tab:ablation} shows ablation study results on the S3DIS dataset. 
First, when replacing the \textit{G-CNN} with a traditional CNN, the mean IoU decreases by $7.79\%$, which indicates that symmetry information 
is useful. 
We further verify the usefulness of RBF-VAE.
We found that without RBF, \{0,1\}-VAE often fails to produce reasonable results. This is because point clouds are sparse in 3D space. For example, in S3DIS, each point cloud contains $4096$ points. Over $64\times 64\times 128$ subvoxels, the average point density per subvoxel is only $0.008$.  In this example, Our original grid size is $(64,64,128)$ and the input \{0, 1\} volume is of size $16MB$. With the RBF-VAE, the input size is reduced to $2MB$. Compared with taking \{0, 1\} subvoxels as input, our RGB-VAE scheme significantly reduces the memory consumption and computational cost. As shown in the table, it further helps improve the performance.

\GL{We replaced our VAE with AE and highlight the performance of this modified approach on the S3DIS dataset in Table~\ref{tab:ae}. The average reconstruction losses for both AE and VAE are close on the \emph{training} set, while average reconstruction loss for AE is about $2.2\times$ higher than that for VAE on the \emph{test} set. The VAE has significantly better generalizability than AE due to the prior distribution assumption and avoids potential overfitting to the training set. }
\begin{table}
    \caption{\textbf{Ablation study on S3DIS dataset.} 
    First row: original results;
    Second row: replacing G-CNN with traditional CNN; Third row: replacing RBF-VAE with RBF grids; Fourth row: replacing RBF-VAE voxels with \{0,1\} grids. 
    Note that the VAE latent variable distribution is designed for incorporation with RBF. We also considered directly applying G-CNN on RBF subvoxels, but that was not useful due to the compact representation of VAE encoding and lowers the performance.
    }
\begin{adjustbox}{max width=0.5\textwidth}

\centering

    \begin{tabular}{llll}
    \hline
    & Overall Acc & mean IoU & mean IoU threshold $0.5$\\
    \hline
    Ori.(G-CNN + $16\times16\times32$ RBF-VAE) & 85.98 & 75.40 & 79.00 \\
    Trad. CNN + $16\times16\times32$ RBF-VAE & 80.67 & 67.61 & 71.43\\
    G-CNN + $32\times32\times64$ RBF &78.15 &64.13 & 68.11\\
    G-CNN + $64\times64\times128$ finer grid& 82.36& 70.00& 74.14\\
    \end{tabular}

    \label{tab:ablation}
    \end{adjustbox}
    \vspace{-10px}
    \end{table}

\begin{table}[]
    \caption{\textbf{Ablation Study on VAE.} 
    First row: our original results; second row: The VAE function is replaced with AE function. The same parameter settings are used ($l=8, k=4$). We observe better accuracy with our original VAE.}
    \label{tab:ae}
    \centering
    \begin{adjustbox}{max width=\linewidth}
    \begin{tabular}{llll}
    \hline
    & Overall Acc. & mean IoU & mean IoU threshold $0.5$\\
    \hline
    VAE (original algorithm) & 85.98 & 75.40 & 79.00 \\
    RBF-AE+GCNN (modified algorithm)  & 82.07 & 69.60 & 73.38\\
    \hline
    \end{tabular}
    \end{adjustbox}
    \vspace{-10pt}
\end{table}

\section{Conclusions, Limitations and Future Work}\label{sec:conc}

In this paper we introduced a novel Voxel VAE network (VV-Net) for robust point segmentation. Our approach uses a radial basis function based variational auto-encoder and combines it with group  convolutions. We have compared its performance with state-of-the-art point segmentation algorithms and demonstrate improved accuracy and robustness on well-known datasets. 
{\color{blue}}
While we observe improved performance in most categories, occasionally our approach may not perform well for some input shapes. As in Figure~\ref{fig:fail}, the network suggests that the \textit{Cap} is a \textit{Table}, which may be caused by the group convolutional module because the module encodes $90^\circ$ symmetry.
As future work, we would like to further improve the accuracy and evaluate the performance on other complex point cloud datasets. 
The VV-Net architecture can also be used for  other point cloud processing tasks such as normal estimation, which we will investigate in future.
\section*{Acknowledgements}\label{sec:ack}
This work was supported by National Natural Science Foundation of China~(No. 61828204 and No. 61872440), Beijing Natural Science Foundation~(No. L182016), CCF-Tencent Open Fund, Youth Innovation Promotion Association CAS and NVIDIA Corporation with the GPU donation. 

{\small
\bibliographystyle{ieee_fullname}
\bibliography{egbib}
}


\end{document}